# The Adaptive Optics System for the Gemini Infrared Multi-Object Spectrograph: Performance Modeling

Uriel Conod[1,2], Kate Jackson[2], Paolo Turri[1,2], Scott Chapman[1,2,3], Olivier Lardière[2] , Masen Lamb[4,5], Carlos Correia[6], Gaetano Sivo[5], Suresh Sivanandam[4,7], and Jean-Pierre Véran[2]

[1] Department of Physics and Astronomy, University of British Columbia, 6224 Agricultural Road, Vancouver, BC, Canada; urielconod@phas.ubc.ca
[2] NRC—Herzberg Astronomy and Astrophysics, 5711 West Saanich Road, Victoria, BC, Canada
[3] Department of Physics and Atmospheric Science, Dalhousie University, Halifax, NS, B3H 4R2, Canada
[4] Dunlap Institute for Astronomy & Astrophysics, 50 St. George Street, Toronto, Ontario, Canada
[5] Gemini Observatory—AURA, Colina El Pino s/n, Casilla 603 La Serena, Chile
[6] Space ODT—Optical Deblurring Technologies, Porto, Portugal
[7] David A. Dunlap Department of Astronomy & Astrophysics, 50 St. George Street, Toronto, Ontario, Canada

## Abstract

The Gemini Infrared Multi-Object Spectrograph (GIRMOS) will be a near-infrared, multi-object, medium spectral resolution, integral field spectrograph (IFS) for Gemini North Telescope, designed to operate behind the future Gemini North Adaptive Optics system (GNAO). In addition to a first ground layer Adaptive Optics (AO) correction in closed loop carried out by GNAO, each of the four GIRMOS IFSs will independently perform additional multi-object AO correction in open loop, resulting in an improved image quality that is critical to achieve top level science requirements. We present the baseline parameters and simulated performance of GIRMOS obtained by modeling both the GNAO and GIRMOS AO systems. The image quality requirement for GIRMOS is that 57% of the energy of an unresolved point-spread function ensquared within a $0.1 \times 0.1$ arcsecond at $2.0\,\mu$m. It was established that GIRMOS will be an order $16 \times 16$ adaptive optics (AO) system after examining the tradeoffs between performance, risks and costs. The ensquared energy requirement will be met in median atmospheric conditions at Maunakea at 30° from zenith.

*Unified Astronomy Thesaurus concepts:* Astronomical instrumentation (799); Astronomical simulations (1857); Telescopes (1689)

## 1. Introduction

Near-infrared, mid-spectral resolution spectrographs with multiple integral fields units (IFUs) deployable over a large field of regard (FoR) and assisted by Multi-Object Adaptive Optics (MOAO), have been recently identified as instruments of major scientific interest. The image quality delivered by the MOAO system is a key parameter to fully take advantage of the telescope aperture, as well as to improve significantly the scientific capabilities of the integral field spectrographs (IFS). While multiplexing enables carrying out large AO surveys that would be prohibitive for single-object AO IFUs, such as an extensive IFU survey of high-$z$ galaxies. In AO, the main limitation for a correction over a large field is the so called *anisoplanatism*, which degrades the AO performance for sources at a slightly different (off-axis) position than the targets used for the wave front sensing (Fried 1982). In order to optimize the image quality of ground-based telescopes over a large field of view (FoV), several AO correction techniques can be considered. All these techniques require the measurement of the turbulence volume above the telescope. For more complete correction, multiple wave front sensors (WFS) coupled with laser guide stars (LGS) and natural guide stars (NGSs) are generally used for this purpose. The utility of LGS is driven by the fact that the sky coverage with bright stars is limited. Consequently, it is not possible to find an NGS bright enough and not too far off-axis (from the science target) for wave front sensing everywhere on the sky. Using a laser to generate artificial stars, it is possible to extend the sky coverage significantly (Foy & Labeyrie 1985; Tallon & Foy 1990). NGSs are however still required for tip-tilt measurements.

To obtain uniform AO performance over a large FoV, the first technique consists in correcting for only the ground layer of turbulence (GLAO, Rigaut 2002). The signal coming from multiple WFSs distributed over the FOV can be simply averaged or the ground layer can be more optimally estimated using a least square (LS) or a minimum mean square error (MMSE) reconstructor (Fusco et al. 2001; Ellerbroek 2002).

Often called *seeing enhancement*, this technique provides a relatively uniform correction over a large FoV, but the performance is strongly dependent on the relative strengths of ground and higher level turbulence and results in performance still far from the diffraction limit, see Madec et al. (2018) and Kolb et al. (2017).

MOAO (Hammer et al. 2002) is essentially driven by multi-object spectroscopy (MOS), which consists in inserting several slits/fibers or IFSs in a large FoV to obtain the spectra of several objects simultaneously. This technique is used for crowded fields with astronomical targets like star clusters or galaxy clusters to study the properties of several individual objects, which are generally extended and faint. Instead of correcting the full FOV, MOAO corrects only the wave front in the direction of science objects. For that, a DM for each science target is placed in the optical train upstream of the injection of the light into the IFS. The challenge in MOAO is that the scientific targets are generally too faint to perform rapid wave front sensing, preventing the operation of the DM in closed-loop. The DM is therefore controlled in open-loop using the tomographic information (Ragazzoni et al. 1999) provided by WFSs placed upstream of the DM. MOAO can in principle operate in addition to any other AO system. It has already been demonstrated in simulation that operating an MOAO system downstream of first a GLAO or Multi-conjugate AO (MCAO) system, (a woofer-tweeter MOAO system) will significantly improve the image quality in the area of the MOAO correction, (Morel et al. 2016; Chapman et al. 2018; Jackson et al. 2019). Note that a two-stage (GLAO+MOAO) correction has already been demonstrated on sky (Gendron et al. 2016) and was the baseline configuration for the phase A design of the future multi-object spectrograph MOSAIC Morris et al. 2018, planned for the E-ELT. However, MOSAIC recently decided to no longer implement MOAO due to the coarser spaxel size of the multi-IFU units, but also because of complexity and costs (Sánchez-Janssen et al. 2020).

### 1.1. MOAO Demonstrators

Because of the risks associated with MOAO such as the calibration and the open-loop control of an AO system, several demonstrators have been deployed in the past years. We present and highlight in this section the issues and the progress made with MOAO instrumentation.

The first generation of MOAO demonstrators, with the Visible Light Laser Guidestar Experiments (ViLLaGEs, Ammons et al. 2008; Gavel et al. 2008) and the Victoria Open Loop Testbed (VOLT, Andersen et al. 2008), were focused on open-loop control. Both ViLLaGes and VOLT performed below expectations at low temporal frequencies and highlighted calibration and alignment issues for open-loop control.

This led to a second generation of MOAO pathfinders such as CANARY at the 4.2 m William Herschel Telescope (Morris et al. 2010; Gendron et al. 2011). CANARY is considered as a pathfinder for MOSAIC, with the goal of performing NGS and LGS based wave front sensing, open-loop control and developing the calibration and alignment techniques. The project started in 2010 with the final phase in 2015 with the demonstration on-sky of a woofer-tweeter MOAO observing mode, (Gendron et al. 2016). The woofer DM ($8 \times 8$ actuators) delivers a GLAO correction based on a four LGSs and three NGSs tomography while the DM tweeter ($17 \times 17$ actuators) performs the additional on-axis MOAO correction. A Strehl ratio (SR) of 20% in H-band was obtained with this configuration under a seeing condition of $1''$ (given at 500 nm). Because the tweeter DM exhibited creeping (estimated at 2% at any timescale), a figure sensor based on a $14 \times 14$ subaperture Shack–Hartmann was used to lock the shape DM. We note that CANARY is still in operation and offers now a limited number of nights for science.

The RAVEN MOAO demonstrator (Lardière et al. 2014) installed on the SUBARU Telescope was the first and only MOAO instrument mounted on a 8 m telescope and feeding an AO-optimized science instrument, the Subaru Infrared Camera and Spectrograph (IRCS, Tokunaga et al. 1998). RAVEN had two science channels, each one corrected by a $11 \times 11$ actuator DM. RAVEN demonstrated on-sky performance estimated at 30% EE in a $0.14''$ slit using one LGS and three NGSs with a system operating at 100 Hz.

Despite several MOAO demonstrators deployed in the past years, there are no MOAO instruments installed on 8 meter-class telescopes in operation and delivering scientific data as a facility. The MOAO technique is now mature enough to be considered in the design of future scientific instruments on large aperture telescopes.

### 1.2. GIRMOS

In response to the scientific needs and as a technology demonstrator for Extremely Large Telescopes (ELTs), the Gemini Infrared Multi-Object Spectrograph (GIRMOS) was proposed as a new facility for the Gemini telescope. GIRMOS will allow simultaneous high-angular-resolution, spatially resolved infrared (1–2.4 $\mu$m) spectroscopy of four objects within a $2'$ FoR using MOAO. Thanks to the multiplexing and the excellent image quality requirements, the survey efficiency of GIRMOS will be significantly improved compared to other IFSs installed on 8 meter-class telescopes such as NIFS (McGregor et al. 2003), OISIRIS (Larkin et al. 2006) or MUSE (Bacon et al. 2010). Thus, making GIRMOS a survey instrument of world-class interest. In addition, GIRMOS will play a significant role for the future development of multi-object spectrograph concepts for the TMT such as IRMOS (Eikenberry et al. 2006)

GIRMOS was initially designed as an instrument to be installed behind the Gemini Multi-Conjugate AO system

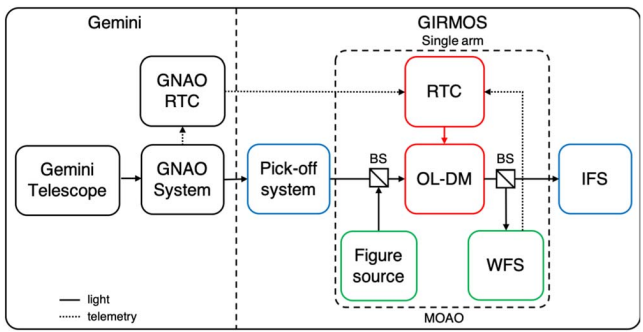

**Figure 1.** GIRMOS functional block-diagram with interfaces between Gemini telescope, GNAO system and GIRMOS.

(GeMS, Rigaut et al. 2014) at Cerro Pacon, Chile (Chapman et al. 2018; Sivanandam et al. 2018). However, during the conceptual phase, several points have been identified in favor of a deployment behind the future Gemini North AO system (GNAO, Sivo et al. 2020), at Maunakea, Hawaii (Sivanandam et al. 2020). First, the better atmospheric conditions at Maunakea would enable a significantly improved AO image quality over Gemini South. Second, the GeMS real time controller (RTC) does not currently have functionality to be able to deliver the telemetry needed to control the GIRMOS open-loop deformable mirrors. Therefore, a major upgrade of the GeMS RTC would have been required to operate the MOAO system of GIRMOS (Chapman et al. 2018).

GNAO is the new AO facility optimized for surveys and time domain science at Gemini North. Equipped with four LGS, GNAO will provide a wide-field mode ($2'$ circular) with GLAO correction and a narrow-field mode ($20 \times 20''$) using Laser Tomography AO (LTAO) correction.

GIRMOS is now designed to operate downstream of GNAO and an near-infrared imager (1–2.4 $\mu$m) is now included in the instrument. To achieve the scientific goals of GIRMOS, an image quality requirement corresponding to 57% ensquared energy (EE) within $0.1 \times 0.1$ arcseconds for a point-spread function (PSF) in HK-band (2.0 $\mu$m) in median atmospheric conditions at Maunakea, was established. Because the HK-band is not generally used by the AO community to extract performance, we use the H-band (1.65 $\mu$m) to measure our metrics for this publication. Assuming the same residual wave front error, we can compute a PSF in different bands (H, HK) and adjust the ensquared energy (EE). We find a difference of 5%, so the image quality requirement in H-band corresponds to 52% EE within $0''\!.1 \times 0''\!.1$. During the conceptual design phase of the instrument, we identified that GNAO would not be able to deliver such an image quality on its own (Sivo et al. 2020), therefore, a MOAO module needs to be included in the GIRMOS object selection mechanism before the injection of the light into the IFS to obtain the necessary image quality.

In Figure 1, we present, the functional principle of GIRMOS and its interaction with GNAO. The MOAO system consists of four main components: A DM controlled in open-loop (OL-DM), a figure source (FS) to illuminate the OL-DM, a Shack–Hartmann WFS that can be used either with the FS or directly with the science target (if compact and bright enough), and finally, an RTC to process the WFS telemetry data and to compute the OL-DM commands. GIRMOS RTC must communicate with GNAO's RTC to obtain the LGS-WFS and NGS-WFS slopes and this could be challenging (interface and synchronisation). However, both RTCs are designed on the same platform (HEART, Dunn et al. 2022; Mueller et al. 2022) and the communication between the two RTC is included in the design through interface control documents between GNAO and GIRMOS.

In this article, we present the performance of the GIRMOS-MOAO system. In Section 2, we describe OOMAO, the simulator used to model the system and assess its performance. In Section 3, we describe the baseline system parameters and performance modeling of GIRMOS-MOAO. In Section 4, we present the exploration of the MOAO parameter space which led us to our baseline design for GIRMOS-MOAO. Section 5 assesses the performance of GIRMOS-MOAO on distant galaxies. Finally, we summarize our results and discuss the future development of the MOAO system for GIRMOS.

## 2. Simulation Tools

### 2.1. OOMAO

Object-Oriented Matlab Adaptive Optics Toolbox (OOMAO, Conan & Correia 2014), is a library of Matlab classes allowing the modeling of AO systems. The numerical simulation consists in the propagation of the wave front through each component of the system, including the sources, the atmosphere, telescopes, DMs, WFSs and imager. OOMAO can simulate NGS and LGS asterisms as well as science targets at different wavelengths, magnitudes and positions in the FoR. The great advantage of using an end-to-end simulation, is the availability at each temporal step, to analyze the wave front at different position in the light path as well as the telemetry of each module (WFS, DM). We note that the performance modeling for RAVEN was also computed with OOMAO and compared to MAOS (a C-based tomographic AO simulator) with excellent agreement (Andersen et al. 2012).

### 2.2. GIRMOS Performance Modeling

The GIRMOS-MOAO system is simulated on top of GNAO operating in GLAO mode. The GNAO system is simulated using the following principle. The atmosphere, the telescope, the ground-layer DM and the wave front sensors are directly associated to the GNAO part of the simulation. The MOAO system consists essentially of a DM operating in open-loop. To command the OL-DM, the telemetry from the GNAO LGS

WFSs is processed. The commands are computed using a spatio-angular minimum-variance tomographic controller as presented in Correia et al. (2015) and used successfully on sky with RAVEN by Lardière et al. (2014).

## 3. Baseline Performance of GIRMOS-MOAO

During, the conceptual design phase of the instrument, a performance and error budget modeling was conducted and presented in Jackson et al. (2019). This work assumed GIRMOS operating downstream of GeMS, and a comparison between the simulation and real data obtained from GeMS in GLAO showed a very good agreement. Even with slightly different parameters (as it is the case for GNAO), we are confident in the modeling of the system and the baseline performance can be trusted. For this publication, we essentially revised this preliminary work according to the expected properties of GNAO and Maunakea atmospheric profiles.

GIRMOS will primarily operate using the integral field spectroscopy (IFS) mode. The main data IFS product will essentially be H$\alpha$ images and the primary driver is the size of clumpy structures in the galaxies. We, therefore, identified that the ensquared energy (EE) within a square spaxel size of $0''\!.1$ is an important metric and will drive the spectral signal to noise ratio. For the imaging side, the Strehl ratio (SR) will give valuable information on the image quality. Because $0''\!.1$ is significantly larger than the diffraction-limited spot ($0''\!.04$) in H-band, the MOAO performance is less sensitive to the low spatial order wave front errors than the high spatial order errors.

### 3.1. Ensquared Energy in $0''\!.1$

The heart of GIRMOS consists of four IFS with three different spaxel sizes ($0''\!.025$, $0''\!.05$ and $0''\!.1$), and an image quality requirement is specified for the $0''\!.1$ spaxel size in HK-band. EE is not a natural metric for use in the design of an AO system. This is essentially due to the fact that no relation exists directly linking the EE with the residual wave front root mean square (rms), as it does for the Strehl ratio. It is therefore difficult to establish a precise error budget as well as to identify the impact of the different error contributors on EE in $0''\!.1$ ($EE_{0.1}$). However, by using an end-to-end simulator such as OOMAO, we can compute the long exposure PSFs and then extract the corresponding EE profiles.

### 3.2. Strehl Ratio

SR is a useful metric to asses the image quality of an optical system, being relative only to the diffraction limit. The SR is defined as the ratio of the AO PSF's central peak intensity with the central peak intensity of the diffraction limited PSF. It is a unit-less metric generally normalized to 1 or expressed as a percentage. The SR is the most common metric used in the AO community and was first introduced by Strehl (1895) and Strehl (1902). This parameter is directly linked by the Maréchal approximation ($SR = \exp(-\left(\frac{2\pi\sigma}{\lambda}\right)^2)$) to another useful metric: the root mean square deviation of the wave front phase $\sigma$. Because GIRMOS has no established SR requirement (as it does for EE) and for the purpose of this paper, we fixed our own threshold value based on the simulation of the system. The idea here is to identify, and apply the offset between the SR and the $EE_{0.1}$ values. Assuming the baseline parameters we simulated H-band PSFs and we estimated a difference of 30%, see Section 3.4.2. We can therefore fix our SR threshold value in H-band at 22%. This EE to SR relation in indeed valid for baseline conditions but can slightly vary depending on the effect considered, as shown as example in Section 3.4.2 Figure 6 for poor seeing or in Section 4.6 Figure 10 for strong NCPAs. However, this SR threshold value remains useful to evaluate performance. Using the Maréchal approximation, this 22% SR corresponds to a wave front error (WFE) deviation of $\sigma = 320$ nm. It is interesting to note that in the case of GIRMOS, our simulations are showing a difference of about 10% between the true SR (directly measured) and the SR computed from the WFE. This difference is probably due to either a large WFE where Maréchal approximation is not valid anymore or a considerable WFE deviations from near-stationarity.

### 3.3. Baseline Parameters

We present the baseline parameters for the GIRMOS-MOAO system. For our simulations we can divide the system essentially in 3 parts. The atmosphere, from which the light sources (LGS and TT stars) pass through and where the turbulence is generated. The GNAO system including the telescope, WFSs (LGS and tip-tilt) and one DM (GNAO-DM0). Finally, the GIRMOS-MOAO system including the OL-DM, a WFS and a FS.

#### 3.3.1. Atmospheric Profiles

The atmospheric parameters are for Maunakea and are adapted from TMT site testing campaign (Els et al. 2009). The baseline profiles are for 50th percentile with 7 layers, see Table 1. The outer scale of the turbulence is fixed to 40 m and does not strongly impact the EE and SR performance.

#### 3.3.2. GNAO–GLAO Parameters

We note that the parameters used for this section are in good agreement with GNAO's preliminary study and modeling presented in Sivo et al. (2020). In our simulations we consider a Gemini primary aperture of 8 m with a central obscuration ratio of 0.14. For all this publication, the telescope is assumed to be pointing at a zenith angle of 30 degrees. The baseline observing mode considered for GNAO is GLAO, using four LGSs and three NGSs located in the $2'$ FoR as presented in Figure 2. Each LGS has a magnitude of $V = 13$ and is coupled to a Shack–Hartmann WFS with $20 \times 20$ subapertures and $8 \times 8$ pixels per lenslet. NGSs with a baseline magnitude of $V = 17$, are respectively coupled to a wave front sensor consisting of a simple imager (SHWFS with one lenslet) with $20 \times 20$ pixels. For both NGS and LGS WFSs we assumed a read-out-noise of $0.5 e^-$ rms. Note that this value can be reached with EMCCD type detectors with the counterpart of the known *excess noise*, which is equivalent to double the photon noise. We simulated the ground layer DM (GNAO-DM0) with $21 \times 21$ actuators. The system controller is an integrator with a loop gain of 0.6. A summary of he GNAO baseline parameters used for the simulation is presented in Table 2.

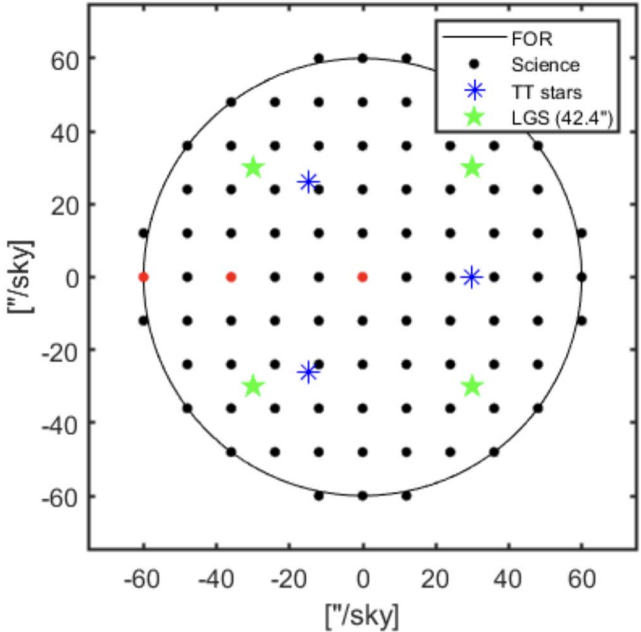

**Figure 2.** Science, LGS and TT stars positions within the GIRMOS FoR. The red dots correspond to positions used to extract GIRMOS-MOAO performance on distant galaxies (see Section 5).

**Table 1**
Turbulence Profile used for Maunakea 25th, 50th and 75th Percentiles with $r_0$ Computed for a Wavelength of 500 nm

| Elevation (m) | Wind speed (m s$^{-1}$) | Wind direction (°) | Turbulence fraction | | |
|---|---|---|---|---|---|
| | | | 25%ile | 50%ile | 75%ile |
| 0 | 5.6 | 190 | 0.5152 | 0.4557 | 0.3952 |
| 500 | 5.77 | 255 | 0.0951 | 0.1295 | 0.1665 |
| 1000 | 6.25 | 270 | 0.0322 | 0.0442 | 0.0703 |
| 2000 | 7.57 | 350 | 0.0262 | 0.0506 | 0.0773 |
| 4000 | 13.31 | 17 | 0.1160 | 0.1167 | 0.0995 |
| 8000 | 19.06 | 29 | 0.0737 | 0.0926 | 0.1069 |
| 16000 | 12.14 | 66 | 0.1416 | 0.1107 | 0.0843 |
| $r_0(m)$ | | | 0.247 | 0.186 | 0.135 |

**Table 2**
Baseline Parameters for the GNAO system

| Telescope | GEMINI NORTH |
|---|---|
| Diameter | 8 m |
| Central Obscuration ratio | 0.14 |
| Zenith angle | 30° |
| **Deformable Mirror** | **GNAO-DM0** |
| Order | $21 \times 21$ |
| Pitch | 0.38 m |
| Coupling | 0.4 |
| Conjugation altitude | 0 m |
| **LGS Wave front Sensor** | **GNAO-LGS-WFS** |
| Number | 4 |
| Order | $20 \times 20$ |
| Pixels per lenslet | $8 \times 8$ |
| Asterism position | 1.′0 square |
| Framerate | 500 Hz |
| Read-out Noise | 0.5 $e^-$ rms |
| $\lambda$ | 589 nm |
| **NGS Wave front Sensor** | **GNAO-NGS-WFS** |
| Number | 3 |
| Order | $1 \times 1$ |
| Pixels per lenslet | $20 \times 20$ |
| Asterism position | 30″ circle |
| Framerate | 500 Hz |
| Read-out Noise | 0.5 $e^-$ rms |
| $\lambda$ | V-band |

### 3.3.3. GIRMOS-MOAO Parameters

For the additional MOAO correction, the system consists only of a DM, a figure source (FS) and a SHWFS. The parameters used for the simulation of these components are shown in Table 3. The baseline order of the OL-DM is $17 \times 17$, see Section 4.1. The WFS ($16 \times 16$) and the Figure Source used to illuminated the OL-DM (a monochromatic source at a wavelength of $\lambda = 780$ nm) are discussed in Section 4.5.

**Table 3**
Baseline Parameters for the GIRMOS-MOAO system

| Open Loop DM | GIRMOS-OLDM |
|---|---|
| Order | 17 × 17 |
| Pitch | 0.5 m |
| Coupling | 0.4 |
| Conjugation altitude | 0 m |
| Open Loop Framerate | 500 Hz |
| **Figure Source** | **GIRMOS-FS** |
| λ | 780 nm |
| **Figure/Truth WFS** | **GIRMOS-WFS** |
| Order | 16 × 16 |
| Pixels per lenslet | 8 × 8 |
| Framerate | 200 Hz |
| Read-out Noise | 2 $e^-$ rms |
| λ | 750–950 nm |

### 3.4. Simulated Performance

The GIRMOS 2′ FoR is sampled by 11 × 11 positions from which we can extract the GLAO and the GLAO+MOAO performance. A representation of the baseline positions of the LGS, NGS and science objects within the 2′ FoR is shown in Figure 2. The simulations were run for a period of 5 s, from which the baseline performance of GLAO and GLAO+MOAO was extracted. This value is a compromise between the computation time and the evolution time of the turbulence.

#### 3.4.1. GNAO (GLAO)

We provide the nominal performance of GNAO operating in GLAO mode. The $EE_{0.1}$ and SR performance are shown in Figure 3 Left. Averaging the values over the 2′ FoR, we find $EE_{0.1} = 51\%$ with a standard deviation across the field of 1% and SR = 11% with a deviation of 1%. For this set of simulations, the ground layer DM commands are computed by averaging the signal of the WFSs (NGS and LGS). The GNAO team is now considering a ground layer tomographic reconstructor optimized over a 85″ square centered in the telescope's FoV. We have implemented in our code this specific mode and obtain as preliminary results a better GLAO correction.

#### 3.4.2. GIRMOS (GLAO+MOAO)

The GIRMOS spectrograph main operating mode is GLAO+MOAO. For this configuration, GNAO is operating as mentioned in the previous section. For this set of simulations, the OL-DM commands are computed using the pseudo-open loop (POL) signal built from the GNAO-LGS-WFSs' residual slopes and the DM0 commands. Because the DM0 is already correcting the ground layer, we need to remove is shape from the OL-DM. However in the baseline configuration, DM0 does not have the same number of actuators than the OL-DM. Using a projection of the GNAO-DM0 shape into the OL-DM actuator space, we can compute the commands needed to remove from the OL-DM. In order to avoid needing to remove the GNAO-DM0 shape, we are now considering the use of only the GNAO-LGS-WFSs' residual slopes. This implies a modification of the reconstructor and different approaches can be considered. Because the ground layer is corrected, the atmosphere is not the same and the reconstructor presented in Correia et al. (2015) has to be modified. Moreover, the ground layer correction is not perfect and the residue is difficult to evaluate. We are currently exploring the impact using an atmosphere with different ground layer residuals of turbulence for the calculation of the optimal reconstructor. A SLODAR scheme as presented in Ono et al. (2017) is considered to obtain the $Cn^2$ profile needed to compute the reconstructor.

The $EE_{0.1}$ and SR performance are shown in Figure 3 Right. Averaging the values over the 2′ FoR, we find $EE_{0.1} = 60\%$ with a standard deviation across the field of 1% and SR = 30% with a deviation of 2% which meet the requirement previously discussed in Sections 1.2 and 3.2.

### 4. Exploring the Simulation Parameter Space

The MOAO parameter space is large and the simulation complex. We therefore identified the most relevant parameters and present their impact on the image quality performance. To reduce the computation time, we decided to explore a set of several values per parameter of interest. For each parameter, we ran a full simulation and we computed a map of the SR and the EE within 0.″1 as shown in Figure 3. When possible, we have linearly interpolated the contours for $EE_{0.1} = 52\%$ for each different parameter values and presented the results using a color code. We also interpolated the contours for SR = 22%. From the interpolated contours we can also compute the fraction of the FoR, where the requirements are meet, as a function of each parameter of interest. Except for the parameters explored, the simulation uses the baseline parameters discussed in the previous section.

#### 4.1. System Order

The system order plays a significant role on the performance of the AO system. It represents the number of WFS subapertures across the pupil diameter and drives the sampling of the wave front. The GNAO system is assumed to be in Fried configuration, the number of DM actuators is therefore equal to the WFS order plus one. Since the conceptual design where the GNAO system order was 16 (Sivo et al. 2020), the system order baseline raised up to 20 (GNAO private communication). So we fixed the GNAO order to 20 and explored the impact of the GIRMOS order only. We noticed that the GNAO order essentially drives the wave front measurement and has a stronger impact than the GIRMOS order. We computed the mean value of SR and $EE_{0.1}$ over the full 2′ circular FOR, for different system order, see Figure 4. We found that the system order does not have a strong impact on the EE performance primarily because of the order of GNAO (20). We did not explore system orders larger than 20 because the wave front sensing order is driven by GNAO and limits the tomographic information. Therefore, exploring an OL-DM with a higher order than the GNAO baseline will not provide better MOAO performance. The baseline order for the GIRMOS-MOAO system is fixed to 17. This number appears to be a good compromise between performance, risks and cost of the hardware, i.e., increasing the number of DM actuators to more than 17 across becomes substantially more expensive and also challenging for the RTC.

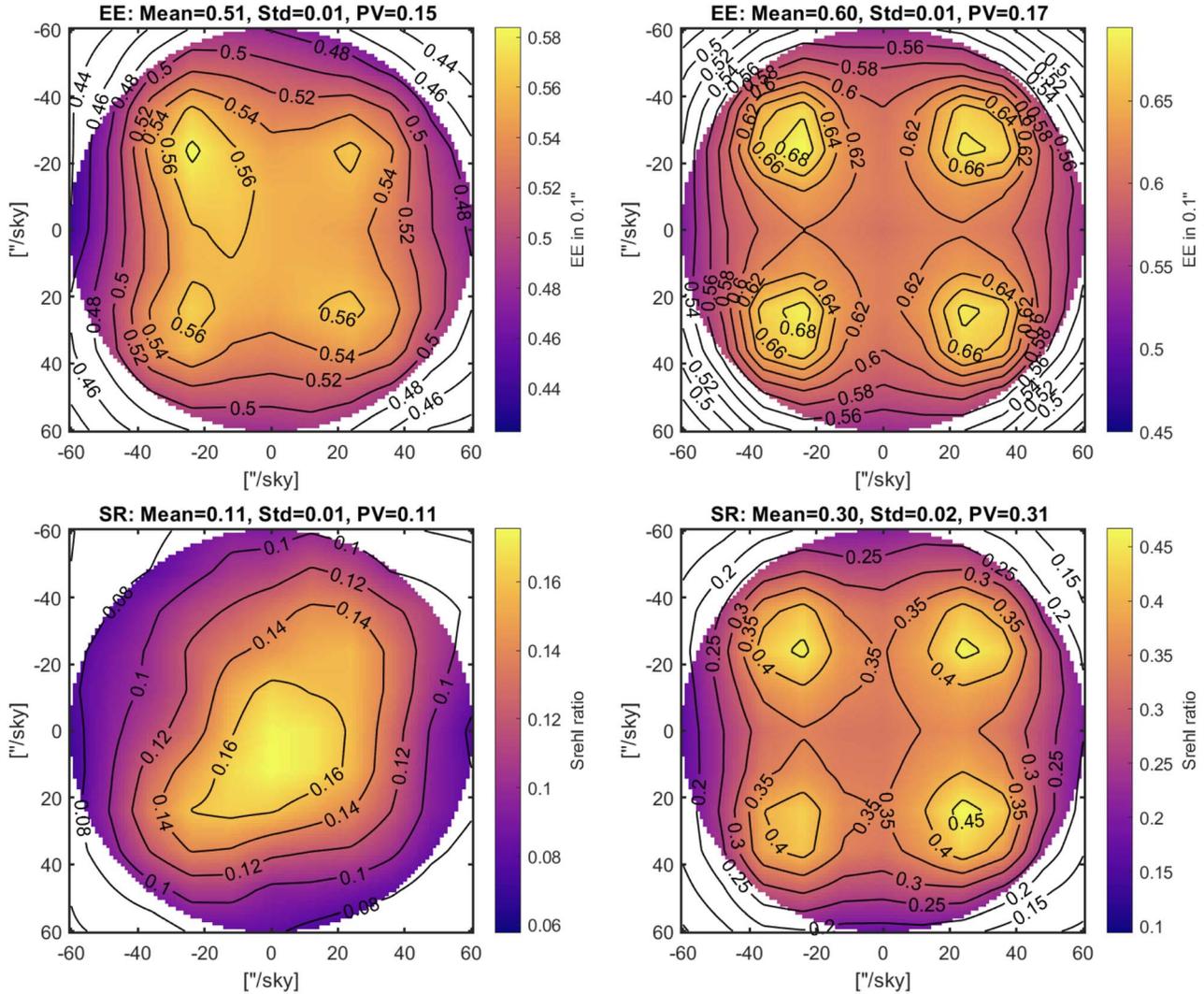

**Figure 3.** Nominal AO performance over the 2′ FoR. H-band $EE_{0.1}$ (top) and SR maps (bottom). Left: GLAO mode with baseline parameters. Right: GLAO+MOAO mode with baseline parameters.

### 4.2. Atmospheric Conditions

We explored different atmospheric conditions essentially through the $r_0$ parameter, while keeping the same 50th percentile atmospheric profile ($Cn^2$). On Figure 5 left, we can see that the area where $EE_{0.1}$ is equal to 52%, reduces drastically for small $r_0$ values. The GLAO as well as the MOAO performance is rapidly limited by the tomographic error as explained in Jackson et al. (2019). From Figure 5 and Figure 6 Right, we find that 25%, 50% and 75% of the FoR is in requirements for respectively $r_0 = 0.138$ m (seeing = 0″.73 at 500 nm), $r_0 = 0.149$ m (seeing = 0″.68) and $r_0 = 0.164$ m (seeing = 0″.62). All these values are between the 50th and 75th percentile of expected $r_0$ values at Gemini telescope. The mean SR value over the FoR is slightly less impacted as the $EE_{0.1}$ for small $r_0$, see Figure 6 Left.

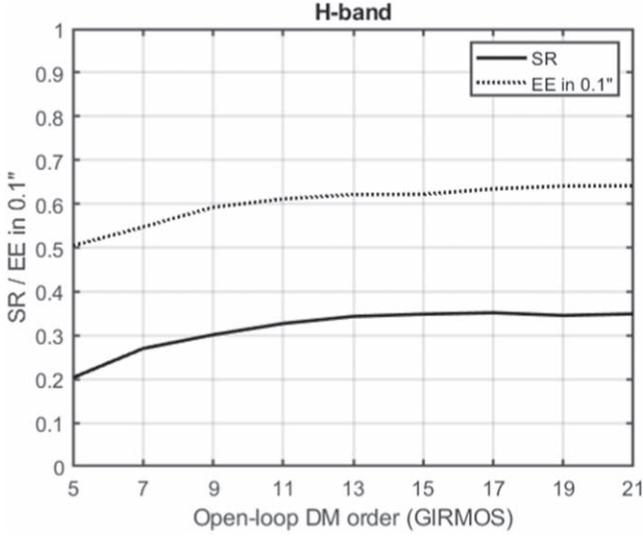

**Figure 4.** GIRMOS open-loop DM order impact, GNAO order is fixed to 20.

GNAO must operate down to a zenith angle of 50° (GNAO private communication). We therefore studied the impact of the zenith angle on the GIRMOS image quality for values from 0° to 60°. The zenith angle $\zeta$ affects the Fried parameter $r_0$ as shown in Equation (1)

$$r_0(\zeta) = \cos(\zeta)^{3/5} r_0(0). \quad (1)$$

From Figure 6 Left, we can extract the zenith angle impact. Using Equation (1) we can compute $r_0(\zeta)$ value assuming the baseline value of $r_0(0) = 0.186$m and a zenith angle of 30°, we find $r_0(30°) = 0.171$. Both the SR and $EE_{0.1}$ averaged over the FoR meet the requirements, and the requirements are met over 25% of the FoR for zenith angle down to 50° which correspond approximately to $r_0 = 0.143$m on Figure 6.

### 4.3. LGS Asterism Position

We explored the impact of different LGS asterism positions across the FoR. The baseline LGS asterism is a square configuration. Therefore, we studied different sizes of the asterism by using three radial distances (50, 45 and 42″.4) for the LGS from the center of the FoR. We found no major impact (<5% relative) on the average EE and SR performance.

### 4.4. DM Requirements

The stroke needed for the OL-DM is an important constraint for the choice of the DM technology and performance. From simulations, we extracted the OL-DM stroke for different seeing ($r_0$), outer scale ($L_0$) conditions as well as different sky positions in the field. Figure 7 Left shows the histogram of the OL-DM actuator stroke for a $r_0 = 0.10$ m (seeing = 1″.0 given at 500 nm), $L_0 = 40$ m for the whole 2′ FOV. We can see that for a seeing of 1″.0 a stroke of 6 $\mu$m will ensure an OL correction after a first stage of GLAO correction, even in very bad atmospheric conditions. However, in order to allow additional corrections such as Non-Common Path Aberrations (NCPAs) and quasi-static aberrations (QSAs), a larger stroke than 6 $\mu$m will be required. We estimated at least an additional stroke of 1 $\mu$m is needed to correct for NCPAs.

### 4.5. Figure/Truth WFS Requirements

We determined the optimal figure WFS FoV requirement in order to parameterize the WFS detector format (number of pixels, pixel scale, lenslet focal-length, lenslet FoV, etc.). We ran OOMAO simulations of GIRMOS (GLAO+MOAO) including a figure source and the figure WFS. We evaluated the cumulative distribution of the figure source photons within each figure WFS sub-aperture, see Figure 7 Right. Only one value, $r_0$ was considered here: 0.10 m corresponding respectively to 1″.0 seeing value (given at 500 nm). We found that for this poor seeing condition, a WFS FoV of 1″.2 should be sufficient to ensure good wave front sensing.

However, this is assuming GNAO providing a standard GLAO correction and 1″.2 is certainly not sufficient to accommodate NCPA offsets or if we want to operated or test GIRMOS on sky without GNAO. To allow wave front sensing in case of large NCPA's offset and strong turbulence, the baseline FoV for the GIRMOS WFS is 4″.4.

### 4.6. NCPA Requirements

We also explored the impact of uncorrected NCPAs on the performance. We simulated random NCPA wave front maps, essentially due to polishing errors using a serie of 100 Zernike modes with their amplitude following a $f^{-2}$ power law. We ran the simulation and add the NCPAs on top each wave front frame used to compute the long-exposure PSF. We repeated the the operation for different amplitudes of NCPAs and the results are shown in Figure 8. We can see that the EE and SR are meeting the requirement for NCPAs amplitudes smaller than 160 nm rms. To mitigate the NCPAs between the OL-WFS and the IFS, we are planing to use a focal plane sharpening algorithm (Lamb et al. 2014), adapted for GIRMOS.

### 4.7. Other Sources of Error

The performance delivered by GIRMOS will be further reduced by additional sources of error, which will impact both the EE and SR. These errors were not previously addressed in individual sections, both because they are difficult to simulate (for example because of a lack of available data to simulate their effects), and they are mostly expected to have insignificant impact. We present in this section, a non-exhaustive list of potential additional errors that should be considered more as the project progresses. The latency of the system has been simulated and the results indicate that no significant difference ($<1\% \, EE_{0.1}$ and $<3\% \, SR$) was found between one or two frame delay. Atmospheric errors, such as chromatic errors, chromatic anisoplanatism, and scintillation, are expected to have a small impact on the performance (Devaney et al. 2008). Other errors, such as alignment and registration errors, as well as the NGS asterism position and the number of NGS available in the field, are currently being explored. Additionally, we have started studying vibrations and possible mitigation strategies.

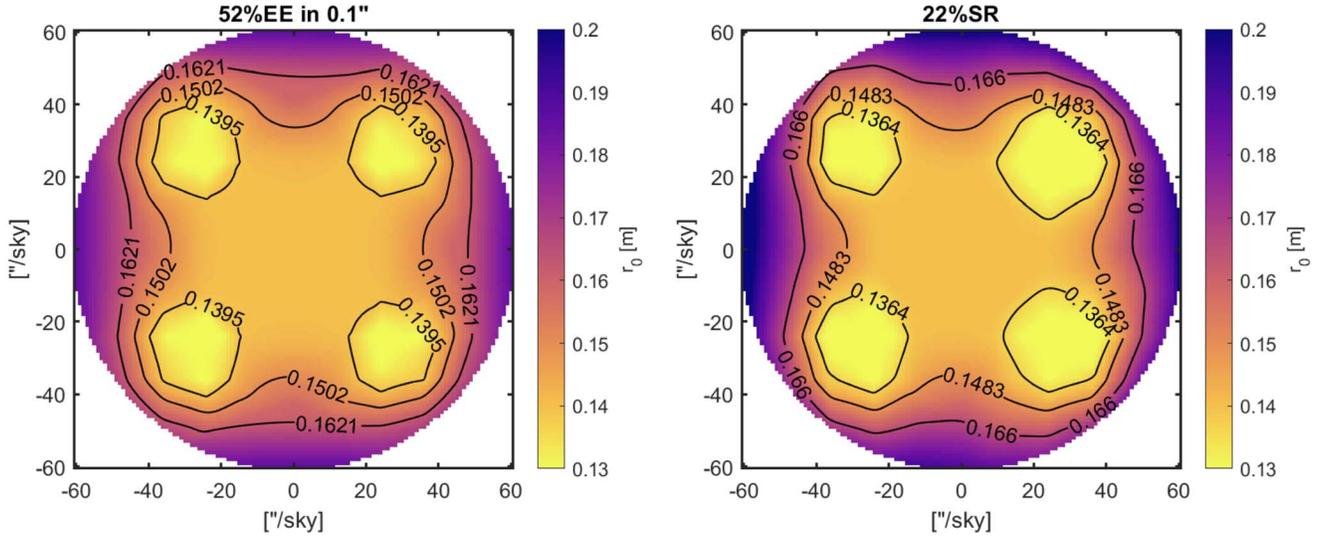

**Figure 5.** $r_0$ values translated at zenith (color code) at which performance threshold is met for each position in the field. Left: $EE_{0.1} = 52\%$. Right: $SR = 22\%$. The three contour lines are for the $r_0$ values for which respectively 25%, 50% and 75% of the FoR is within requirements.

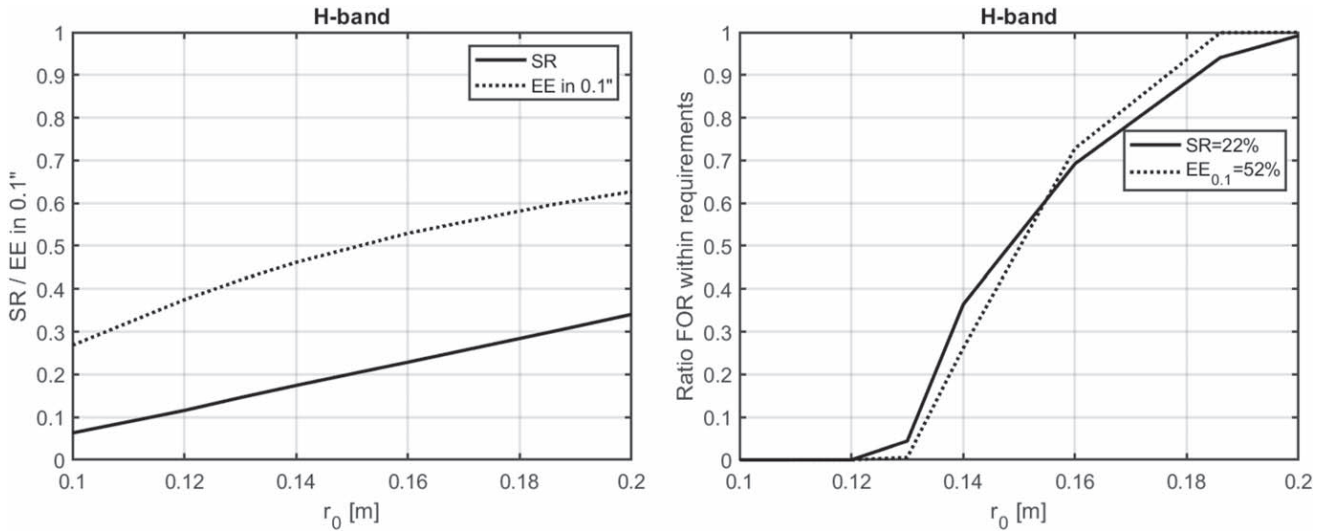

**Figure 6.** *Left*: $EE_{0.1}$ and SR averaged over the $2'$ FoR for different $r_0$ values. Right: Ratio of the GIRMOS FoR where the requirement are reached for both $EE_{0.1} = 52\%$ and $SR = 22\%$.

Another error to mention is the Gemini North M2 print-through which will affect the SR to about 4% in H-band while only having a minimal impact on $EE_{0.1}$ of few percents. However, this known issue should be fixed by the time GIRMOS will be installed at the telescope.

## 5. Simulated Performance of GIRMOS-MOAO on Distant Galaxies

In this section, we apply our simulations to demonstrate the performance of GIRMOS on realistic astronomical targets under a range of atmospheric conditions. A cardinal science case of GIRMOS will be surveys of large numbers of high redshift galaxies. The $z = 2$–2.5 range represents an optimal regime where key diagnostics, H$\alpha$, [NII], and [OIII], H$\beta$, lie in the K- and H-bands, respectively. We thus seek representative galaxy templates to use as a basis for assessing the GIRMOS performance.

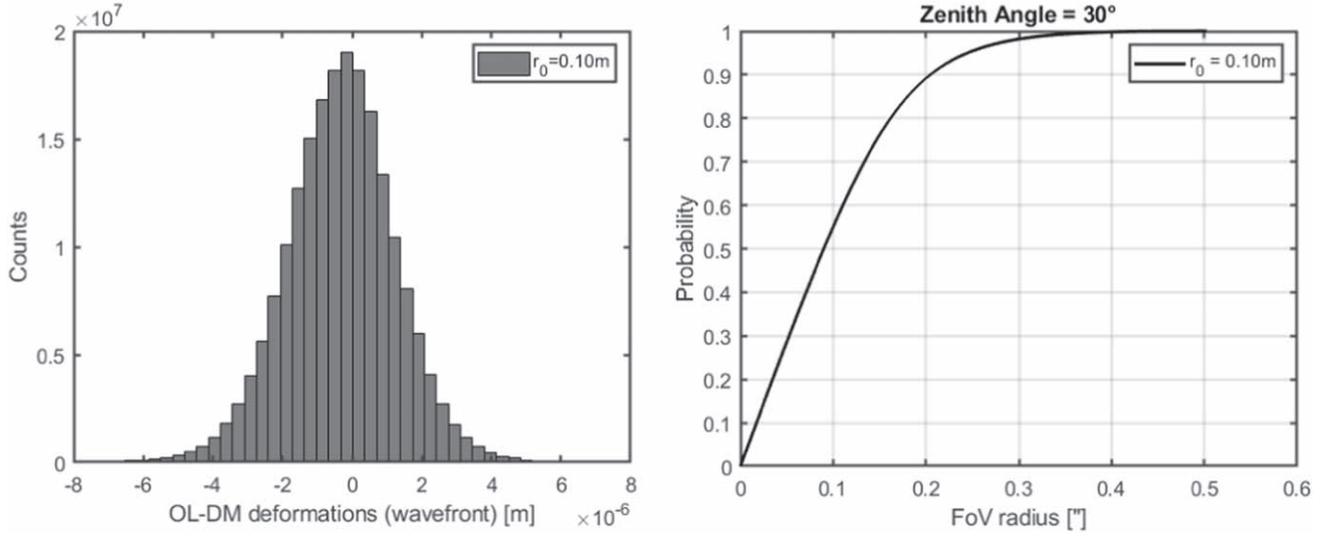

**Figure 7.** *Left*: Histogram of GIRMOS OL-DM deformations recorded from simulations and given in wave front space. Right: Cumulative distribution of wave front slopes as a function of field radius for a turbulence strength corresponding to $r_0 = 10$ cm.

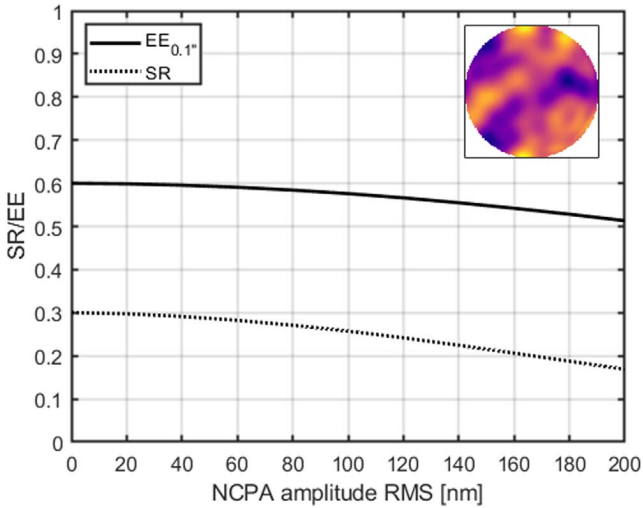

**Figure 8.** $EE_{0.1}$ and SR for different NCPAs (insert) amplitudes, in H-band.

Förster Schreiber et al. (2018) describe the "SINS/zC-SINF AO survey" of 35 star-forming galaxies with a median $z = 2.24$. The observations were performed with SINFONI (Eisenhauer et al. 2003; Bonnet et al. 2004) on the Very Large Telescope (VLT) UT4 telescope. The AO correction by the MACAO module (Bonnet et al. 2003) was performed in Natural Guide Star (NGS) or in Laser Guide Star (LGS) mode.

This survey represents the largest sample with deep AO-assisted near-infrared integral field spectroscopy at $z \sim 2$, and is similar to planned surveys with GIRMOS targeting more than 10 times larger sample. The SINFONI observation, resolve the H$\alpha$ and [NII] line emission and kinematics on scales estimated at $\sim$1.5 kpc from reference PSF stars achieving FWHM in the K-band of $\sim$0.2″.

For our simulations, we select a representative galaxy from this survey, called K20-ID7 at $z = 2.224$, where the reference PSF star showed a FWHM in the K-band of $\sim$0.15″. This galaxy displays several well defined star-forming clumps in the H$\alpha$ moment-0 map (intensity field), as shown in Figure 10, and is ideal to test the varying performance expected from GIRMOS. However, since our goal is to assess the effect of the GIRMOS PSF on detectability of structures in the galaxy, we choose to magnify and resample the K20-ID7 data by a factor of two, resulting in an effective PSF $\sim$0.075″, closer to the diffraction limit of Gemini at K-band (0.″06). At this magnification, the galaxy still remains representative of structure found in more compact galaxies from the Förster Schreiber et al. (2018) survey, while allowing us to apply the GIRMOS simulated PSF on realistic, marginally resolved structures.

Figure 9 shows a direct comparison of the GLAO only versus GLAO-MOAO performance for the same observing conditions. Even if the GLAO only performance is close to the EE requirement of GIRMOS (see Section 3.4.1), there is a substantial improvement in the Strehl ratio, the achieved resolution, and consequently the detection and SNR of clumpy structures using GLAO-MOAO.

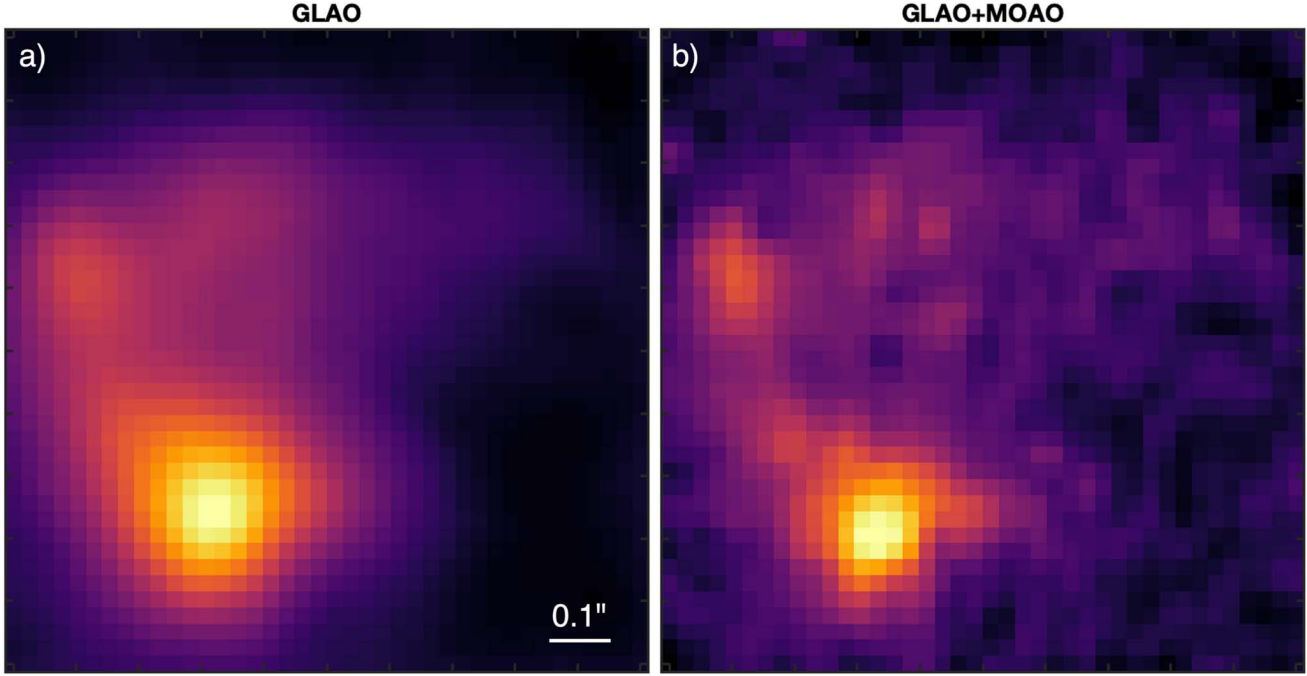

**Figure 9.** Simulated performance of GIRMOS-MOAO on the Hα intensity map of a $z = 2.224$ galaxy, K20-ID7 (Förster Schreiber et al. 2018), observed by VLT-SINFONI with LGS AO. The galaxy has first been magnified by a factor of two before applying the simulated GIRMOS PSF for a $r_0 = 0.20$m and observed at the the center of the FoV. Panel (*a*) is with GLAO only and panel (*b*) is with GLAO+MOAO. The color cut has been optimized independently for each panel to maximize the contrast.

Figure 10 shows nine different GIRMOS simulated PSFs convolved with the magnified Hα intensity map of the K20-ID7, with $r_0 = 0.10, 0.18$, and $0.20$ m, and with positions in the $2'$ field of regard corresponding to $(0'', 0'')$, $(-36'', 0'')$, and $(-60'', 0'')$. The best four combinations in this parameter space are able to reasonably recover the intrinsic properties of the clumpy structure in K20-ID7 (bottom left four panels in Figure 10) through deconvolution of the map with the synthetic (or reconstructed) PSF. In the worst five combinations, the intrinsic sizes and shapes are mostly lost, and the ability to even distinguish the five clumps is progressively worse. However, in all but the worst combination (poor seeing, and edge of FoR), useful resolved properties of the galaxy are recovered. This broadly suggests that two "modes" should be considered for GIRMOS surveys to take advantage of a range of seeing conditions coupled with placements of targets in the FoR: a detailed internal structural assessment, and a basic census of structure. The break point between the two classes is roughly the median seeing at Maunakea ($0\rlap{.}''55$ at R-band, and target locations within approximately $40''$ radius in the FoR.

In Table 4 the analysis of the clumps is quantified by assessing the signal to noise ratio measured for each of the five clumps identified in Figure 10, for each position and $r_0$ value. The angular separation between pairs of clumps before convolution of the GIRMOS PSF is significantly larger than the diffraction limit in all cases. The degrading GIRMOS PSF leads to a progressive loss in measured SNR, and eventually blends clumps A B and C D, as observed qualitatively in the above assessment.

**Table 4**
Reported Signal to Noise Ratio (SNR) values for the Five Different Clumps (A, B, C, D, E) Identified in Figure 10

| | | Clump SNR | | | | |
|---|---|---|---|---|---|---|
| Position | $r_0$ [m] | A | B | C | D | E |
| $(0'', 0'')$ | 0.20 | 19.3 | 8.0 | 9.9 | 6.4 | 6.0 |
| $(-36'', 0'')$ | 0.20 | 17.6 | 7.8 | 8.9 | 6.0 | 5.7 |
| $(-60'', 0'')$ | 0.20 | 15.5 | * | 7.9 | 5.3 | 5.0 |
| $(0'', 0'')$ | 0.18 | 16.2 | 6.9 | 8.3 | 5.4 | 5.0 |
| $(-36'', 0'')$ | 0.18 | 15.3 | 7.1 | 7.7 | 5.2 | 4.9 |
| $(-60'', 0'')$ | 0.18 | 14.7 | * | 7.5 | 5.2 | 4.8 |
| $(0'', 0'')$ | 0.10 | 13.6 | * | 7.0 | 4.8 | 4.2 |
| $(-36'', 0'')$ | 0.10 | 13.0 | * | 6.9 | 5.1 | * |
| $(-60'', 0'')$ | 0.10 | 9.2 | * | * | * | * |

**Note.** The * symbol indicates that the clump is blended with the neighbor (B getting swallowed by A, and eventually E swallowed by D).

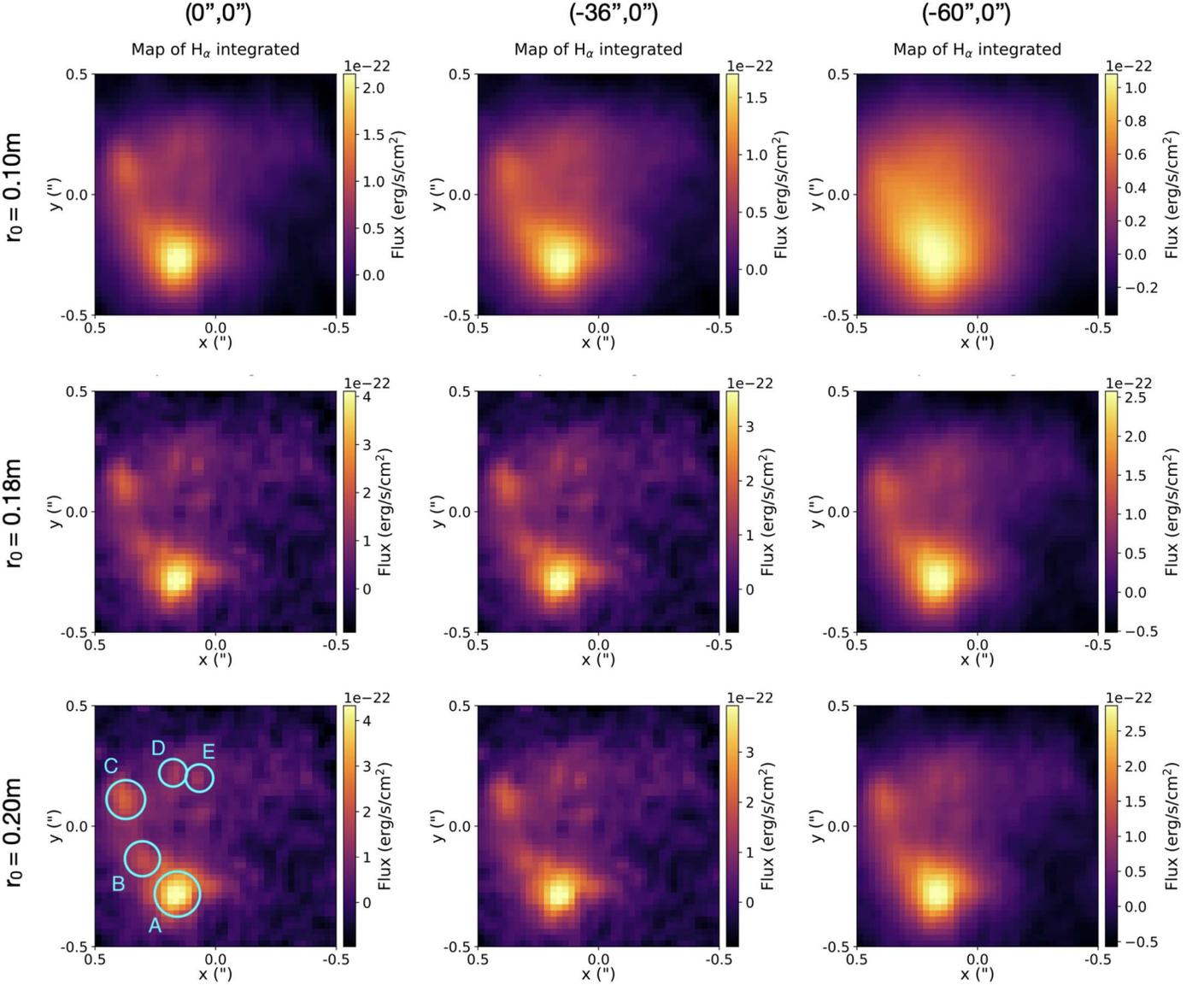

**Figure 10.** Same as Figure 9 but only for GIRMOS-MOAO performance. Nine different GIRMOS simulated PSFs convolved with the magnified H$\alpha$ intensity map of K20-ID7, with seeing characterized by $r_0 = 0.10$, 0.18, and 0.20 m, and with positions in the $2'$ field of regard corresponding to $(0'', 0'')$, $(-36'', 0'')$, and $(-60'', 0'')$, where labels correspond to chosen positions in Figure 2. The five star-forming clump structures exceeding a SNR = 5 are identified in the bottom left image, which can be traced through the degrading PSF in the other panels.

## 6. Summary and Future Developments

A baseline architecture and performance modeling for the MOAO system of GIRMOS has been established. Using OOMAO we simulated both GNAO and the GIRMOS-MOAO system. The observing mode studied in this paper is GNAO operating in GLAO and GIRMOS in MOAO. From simulations, we explored the MOAO parameters space and converged to a system made of a DM with $17 \times 17$ actuators and a WFS with $16 \times 16$ subapertures running synchronously with GNAO up to 500 Hz. The GIRMOS MOAO image quality requirements are meet in median seeing condition. In addition, we demonstrated the performance of GIRMOS on high redshift galaxies for different seeing conditions and positions within the $2'$ FoR.

In order to validate the system performance, we developed a prototype of one GIRMOS-MOAO arm from which encouraging results were found (Conod et al. 2020, 2022). In particular, very low DM open-loop go-to errors were measured for the residual wave front correction, and this *open loop* error was found to scale with increasing amplitude of the residual wave front error. These results validate and drive our design choice of open loop MOAO following a closed loop GLAO correction. The preliminary design phase for the instrument was recently completed and we are now in the critical design phase.

## Acknowledgments

The authors and the GIRMOS project gratefully acknowledges its financial support from the Canada Foundation for Innovation (CFI), Ontario Research Fund (ORF), British Columbia Knowledge Development Fund (BCKDF), Fonds de Recherche du Quebec (FRQ), Nova Scotia Research and Innovation Trust (NSRIT), University of Toronto and in-kind contributions from the National Research Council Canada (NRC) and the Association of Universities for Research in Astronomy (AURA) through its Gemini Observatory.

## ORCID iDs

Olivier Lardière  https://orcid.org/0000-0001-8859-439X